\documentclass[twocolumn]{aa}

\usepackage{graphicx}
\usepackage{txfonts}
\usepackage{xcolor}
\usepackage{colortbl}
\usepackage{hyperref}
\usepackage{array}
\usepackage{arydshln}
\usepackage{comment}
\usepackage[normalem]{ulem} 
\usepackage{multirow,bigdelim}
\usepackage{amsmath,bm}
\usepackage[flushleft]{threeparttable}
\usepackage{rotating}

\newcommand{\bu}{\mathbf{u}}

\usepackage{lastpage}
\makeatletter 
\renewcommand*\aa@pageof{, page \thepage{} of 
\pageref*{LastPage}}
\makeatother

\newcommand{\ii}{{\rm i}}

\newcommand{\bB}{\mathbf{B}}
\newcommand{\bv}{\mathbf{v}}
\newcommand{\br}{\mathbf{r}}

\newcommand{\Blos}{B_{\rm los}}

\newcommand{\Bavg}{B_{\rm avg}^+}

\newcommand{\Bband}{\langle \Bavg \rangle}

\newcommand{\Vlos}{V_{\rm los}}

\newcommand{\Vzon}{V_{\rm zonal}}
\newcommand{\Vavg}{V_{\rm zonal}^-}
\newcommand{\Vband}{\langle \Vavg \rangle}

\defcitealias{2021gizon_inertialmodes}{G+21}

\defcitealias{2024Liang_vlosm1}{LG25}

\makeatletter
\let\linenumbers\relax
\let\endlinenumbers\relax
\makeatother

\begin{document}

\title{Oscillations of the solar photospheric magnetic field\\ caused by the $m=1$ high-latitude inertial mode}

\author{
Stephan~G.~Heinemann\inst{1,2, 3, \thanks{Corresponding author: stephan.heinemann@hmail.at}}
\and Zhi-Chao~Liang\inst{1}
\and Laurent~Gizon\inst{1,4,5, \thanks{Corresponding author: gizon@mps.mpg.de}}
}

\authorrunning{S.G. Heinemann et al.}
\titlerunning{$m=1$ mode in $\Blos$}


\institute{Max-Planck-Institut f\"ur Sonnensystemforschung,  37077~G\"ottingen, Germany 
\and    Department of Physics, University of Helsinki, 00014~Helsinki, Finland
 \and 
Institute of Physics, University of Graz, Universitätsplatz 5, 8010 Graz, Austria   
    \and
 Institut f\"ur Astrophysik und Geophysik, Georg-August-Universit\"at G\"ottingen,  37077~G\"ottingen, Germany
            \and Center for Astrophysics and Space Science, NYUAD Institute, New York University Abu Dhabi, Abu Dhabi, UAE
}
   \date{Received $\langle$date$\rangle$ / Accepted $\langle$date$\rangle$}
\abstract{
Periodic oscillations at 338 nHz in the Earth frame are observed at high latitudes in direct Doppler velocity measurements. These oscillations correspond to the $m=1$ high-latitude global  mode of inertial oscillation. In this study, we investigate the signature of this mode in the photospheric magnetic field using long-term series of line-of-sight magnetograms from the Helioseismic and Magnetic Imager (HMI) and the Global Oscillation Network Group (GONG). Through direct observations and spectral analysis, we detect periodic magnetic field oscillations at high latitudes ($65^\circ$--$70^\circ$) with a frequency of 338 nHz in the Earth frame, matching the known frequency of the $m = 1$ high-latitude inertial mode. The observed line-of-sight magnetic field oscillations are predominantly symmetric across the equator. We find a peak magnetic oscillation amplitude of up to $0.2$~gauss and a distinct spatial pattern, both consistent with simplified model calculations in which the radial component of the magnetic field is advected by the mode's horizontal flow field.
}

\keywords{Sun: magnetic fields --  Sun: oscillations --  
Sun: interior --
Sun: photosphere --
Sun: activity --
Sun: rotation}

   \maketitle

\section{Introduction}

The Sun supports periodic and quasi-periodic oscillations over a wide range of spatial and temporal scales. In addition to the well-known five-minute acoustic modes of oscillation, it also exhibits quasi-toroidal modes in the inertial frequency range, with frequencies comparable to the solar rotation frequency. Solar equatorial Rossby modes were first detected by \cite{Loeptien2018} and later confirmed by \cite{Liang2019} and \cite{Hanasoge2019}. A rich spectrum of additional inertial modes, all retrograde in the Carrington frame, were  identified in  frequency-latitude space by \cite{2021gizon_inertialmodes}. Among these modes, the mode with the largest amplitude is an $m=1$ high-latitude mode with north-south symmetric radial vorticity. The amplitude of this mode can be as high as $20$ m/s at times. It is believed to be baroclinically-unstable \citep{Bekki2022} and saturates via a nonlinear interaction with the Sun's latitudinal differential rotation \citep{Bekki2024}. The velocity features associated with this mode had been noticed at high latitudes by various authors, however it was not then  identified as a global mode of oscillation \citep{1993Ulrich,2001Ulrich,Hathaway2013,Bogart2015}. For a recent review of solar inertial modes, we refer to \citet{Gizon2024}.

The high-latitude  modes were initially identified in long time series of  near-surface flows in the longitudinal ($u_{\phi}$) and colatitudinal ($u_{\theta}$) directions. In particular, modes with $m=1$ and both north-south symmetries were observed within 3~nHz of each other  \citep{2021gizon_inertialmodes}. The mode with the significantly larger amplitude is anti-symmetric in $u_\phi$ (symmetric in radial vorticity) and has a frequency   $\nu_{\rm HL1}^{\rm Carr} = -86.3 \pm 1.6$~nHz in the Carrington frame, with a linewidth of $7.8 \pm 0.2$~nHz and a mean amplitude of $9.8$~m/s over the period 2010--2020 \citep{2021gizon_inertialmodes}. The latitude at maximum amplitude was determined to be close to $67.5^{\circ}$. In a recent paper, \citet[][hereafter \citetalias{2024Liang_vlosm1}]{2024Liang_vlosm1} measured  the mode amplitude in direct Doppler data and found that the mode has remained visible above $60^\circ$ latitude throughout the last five solar cycles since 1967. \citetalias{2024Liang_vlosm1} reported that the mode amplitude exhibits a negative correlation of $-0.50$ with the sunspot number and a strong negative correlation of $-0.82$ with the differential rotation rate near the mode's critical latitude (the latitude at which its phase speed equals the rotational velocity).

In the present paper, we show that the $m=1$ high-latitude inertial mode is detectable in the  line-of-sight (LOS) photospheric magnetic field, and we study the temporal evolution of its amplitude over a period of 17 years ($2007$ to $2024$). In Sect.~\ref{sect:obs} we  describe the datasets used. We detect and charactewrize the magnetic field oscillations  in Sect.~\ref{sect:m1} and compare them  to the  velocity oscillations in Sect.~\ref{sect:comp}. The results are discussed in Sect.~\ref{sect:disc}.

\section{Observational datasets} \label{sect:obs}
To investigate oscillations in the photospheric LOS magnetic field ($\Blos$) in the inertial frequency range, we analyze two long-term datasets. We use 720-s cadence LOS magnetograms taken by the Helioseismic and Magnetic Imager \citep[HMI;][]{2012schou_HMI} onboard the Solar Dynamics Observatory \citep[SDO;][]{2012pesnell_SDO} over a time period of more than 14 years (2010--2024), as well as daily merged magnetograms from the Global Oscillation Network Group \citep[GONG;][]{1996Harvey} spanning 17 years (2007--2024). We compile the datasets at a one-day cadence by computing daily averages when multiple images are available each day. 
The duty cycle of the daily averages is nearly $100\%$ for both HMI and GONG datasets. The original magnetograms have {$4096 \times 4096$} pixels for HMI and {$839 \times 839$} pixels for GONG. Unlike \citetalias{2024Liang_vlosm1}, we did not use the Mount Wilson data because its signal-to-noise ratio is significantly lower. 

The data reduction procedure, which is the same for both datasets, is as follows. Daily magnetograms are binned down to $256\times256$ pixels and remapped onto a uniform grid in Stonyhurst longitude ($\phi$) and latitude ($\lambda$) with a grid spacing of $1^{\circ}$ in both coordinates.
We then compute the latitudinally symmetric component of the magnetic field as
\begin{equation}
    \Blos^+ (\lambda,\phi,t) = \frac{1}{2} \left[  \Blos(\lambda,\phi,t) + \Blos(-\lambda,\phi,t) \right].
\end{equation}
We also computed the anti-symmetric component of $\Blos$, but found no significant signal in the neighborhood of the $m=1$ mode frequency.

To further increase the signal-to-noise ratio, 
we consider averages of  $\Blos^+$ in longitude at every time step:
\begin{equation} \label{eq:bavg}
\Bavg (\lambda, t)=  \frac{1}{N_\phi}
\sum_{|\phi|\le 30^\circ}  \Blos^+ (\lambda,\phi, t) ,
\end{equation}
where $N_\phi=61$ is the number of longitude bins in the sum.
To detect a perturbation in the magnetic field that may be associated with the high-latitude mode, we investigate the evolution of the magnetic field  in the latitude range where the mode's amplitude is most prominent \citep[around or beyond $67.5^{\circ}$, see][]{2021gizon_inertialmodes}. We define latitudinal average of $\Bavg$ over the latitude band $65^\circ \le \lambda \le 70^\circ$:
\begin{equation}
\Bband(t) = \frac{1}{N_\lambda} \sum_{65^\circ \le \lambda \le 70^\circ} \Bavg (\lambda, t) ,
\end{equation}
where the $N_\lambda=6$  is the number of latitude bins in the sum.

To compare with the mode characteristics derived from the reduced LOS Doppler velocity ($\Vlos$), we also compute the following quantities from the HMI and GONG Dopplergrams (see \citetalias{2024Liang_vlosm1} for details):
\begin{align}
    \Vzon(\lambda, t) &=  \frac{1}{\sum_{|\phi|\le 30^\circ} |\sin \phi|} \sum_{\;\;\,|\phi|\le 30^\circ}  (\sin \phi)\ \Vlos (\lambda,\phi, t),\label{eq:vzonal} \\
    \Vavg (\lambda,t) &= \frac{1}{2} \left[  \Vzon(\lambda,t) - \Vzon(-\lambda,t) \right], \\
    \Vband(t) &= \frac{1}{N_\lambda} \sum_{65^\circ \le \lambda \le 70^\circ} \Vavg (\lambda, t).\label{eq:vband}
\end{align}
The zonal velocity, $\Vzon$, serves as a proxy for the longitudinal component of surface velocity \citep{2001Ulrich}.

\section{Detection of the $m=1$ mode in magnetograms} \label{sect:m1}

\begin{figure}
\centering \includegraphics[width=1\linewidth,angle=0]{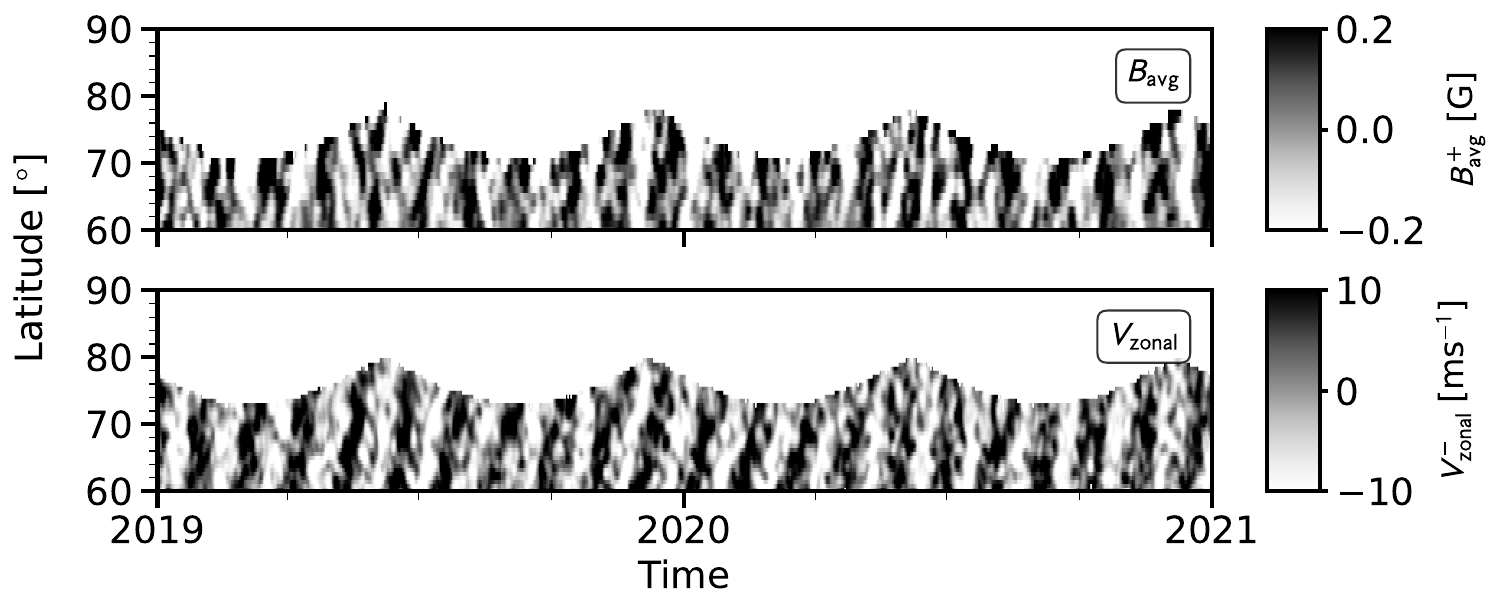}
\caption{Supersynoptic maps in the Earth frame of \(\Bavg\)  and \(\Vavg\) at high latitudes, computed from HMI data for the period 2019--2020. The \(m=1\) mode manifests itself as a series of stripes in both observables. A 180-day running average was subtracted, and for clarity, the maps were smoothed using a Gaussian kernel with a width of two pixels in both latitude and time.}\label{fig:supersynoptic_HL}
\end{figure}

\begin{figure}
\centering \includegraphics[width=1\linewidth,angle=0]{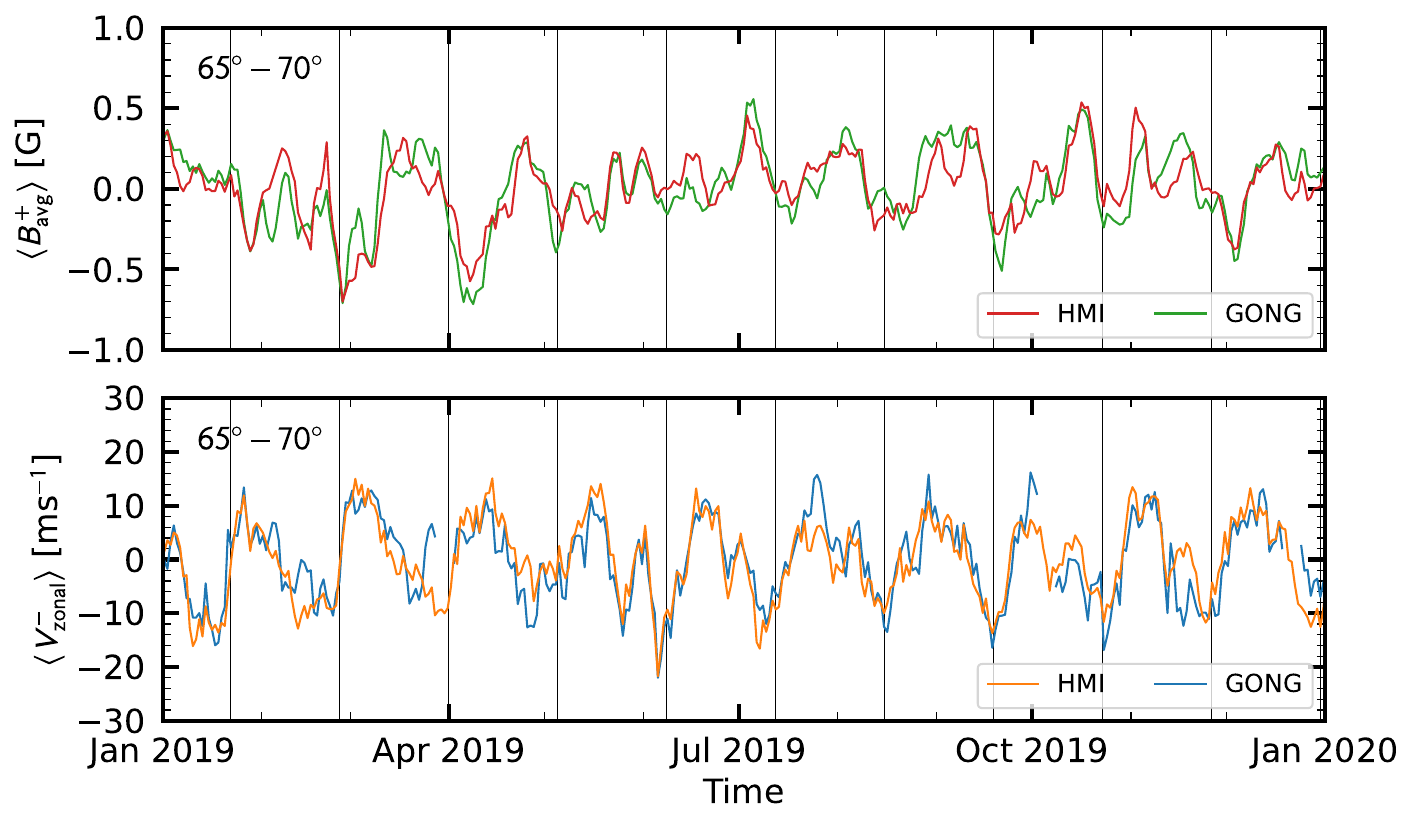}
\caption{Oscillations of $\Bband$  and $\Vband$ in the time series  from HMI and GONG data in 2019. A 180-day average was removed, and for clarity, the maps were smoothed with a Gaussian kernel with $\sigma=1$~day. Vertical lines are spaced at intervals of 34 days.
}\label{fig:real_dom}
\end{figure}

Figure~\ref{fig:supersynoptic_HL} shows supersynoptic maps of  $\Bavg$ and  $\Vavg$ at high latitudes. An oscillating pattern  in both observables is evident in both quantities with a temporal cadence of approximately 34~days. The oscillation is also clearly visible in the spatially averaged data $\Bband$ and $\Vband$ in Fig.~\ref{fig:real_dom}. We find an amplitude of up to $0.5$~G for $\Bband$, associated with the already known amplitude of 10--20~m/s for $\Vband$ in the latitude range $65^{\circ}$--$70^{\circ}$. Since the magnetic field evolution appears to be tightly related to that of the $m=1$ mode observed in the Doppler data \citepalias{2024Liang_vlosm1}, this strongly suggests the presence of the $m=1$ mode in the magnetograms; however, this still needs to be verified through spectral analysis.

\subsection{Power spectra}

\begin{figure}
\centering \includegraphics[width=1\linewidth,angle=0]{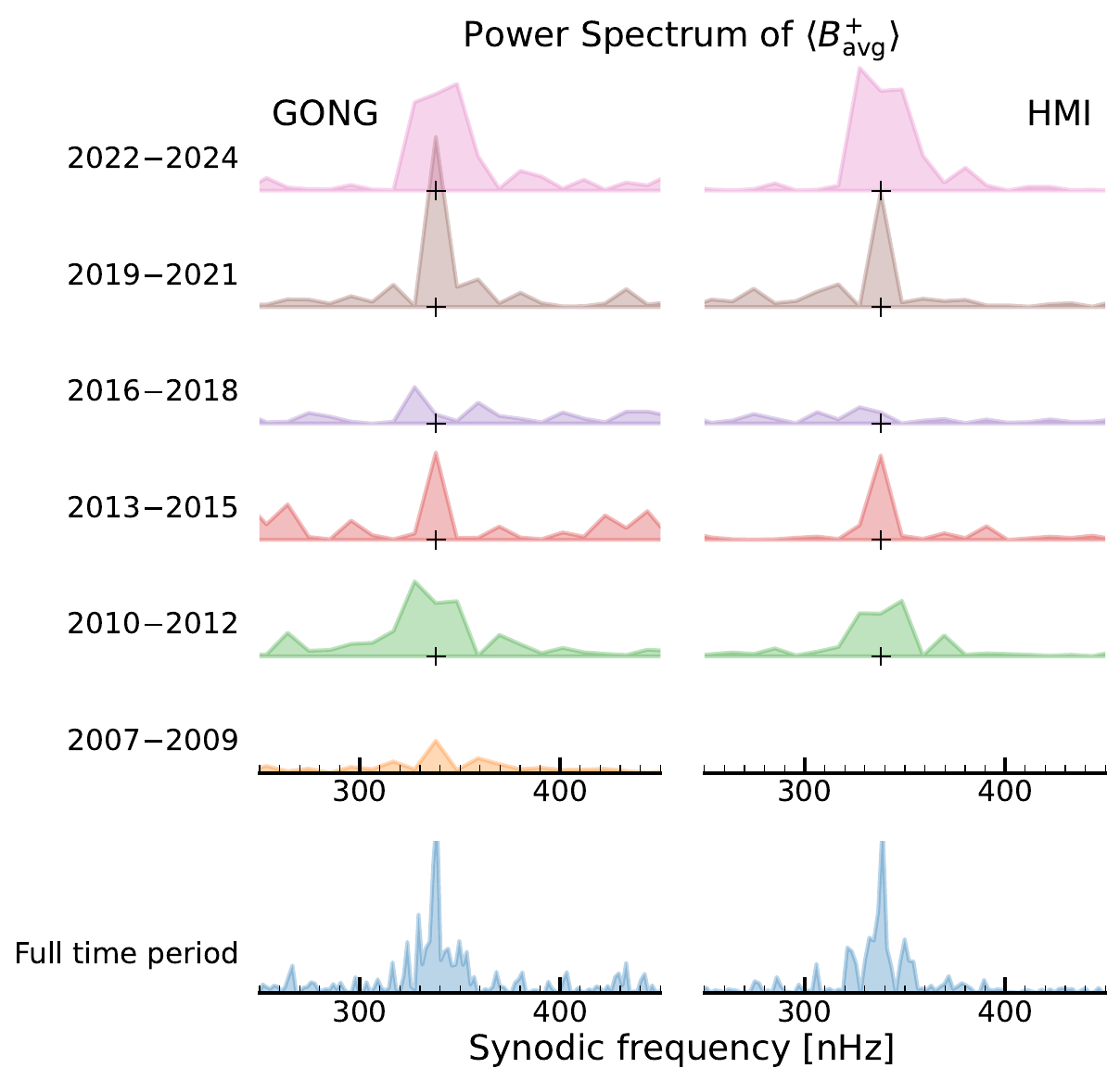}
\caption{Power spectra of magnetic fluctuations $\Bband$ over three-year intervals with a frequency resolution of $10.5$ nHz for GONG (left) and HMI (right) data, stacked vertically and spanning from 2007 to 2024. The crosses mark the reference frequency of 338~nHz. The right-hand-side panel for 2007--2009 contains no data because the HMI dataset starts in 2010. The bottom spectra are computed using the entire available time span for each respective dataset.}\label{fig:four_spectrum_dt}
\end{figure}

We examine  $\Bband$ in the frequency domain by performing a Fourier transform over the full time period available in each dataset. The power spectra were rescaled to account for the missing data in the time series. In  the bottom row of Fig.~\ref{fig:four_spectrum_dt}, we show the power spectra of $\Bband$ for the full HMI and GONG datasets.  In both spectra, we find strong excess power around 338~nHz, with a full width at half maximum of approximately {$11$--$16$~nHz} (corresponding to an e-folding lifetime of $\approx 8$--$11$~months).
We recall that the frequency of the  $m=1$ mode was reported by \citep{2021gizon_inertialmodes} to be  $\nu_{\rm HL1}^{\rm Carr}=-86.3$~nHz in the Carrington frame.
In the Earth frame the mode frequency is 
\begin{equation}
    \nu_{\rm HL1}^{\rm synodic}=\nu_{\rm HL1}^{\rm Carr}+ m(\Omega_{\rm Carr}-\Omega_{\oplus})/2\pi=338~\text{nHz},
\end{equation}
where $\Omega_{\rm Carr}/2\pi =456$~nHz is the Carrington rotation rate and $\Omega_{\oplus}/2\pi = 31.7$~nHz is the Earth's mean orbital frequency around the Sun. 

Figure~\ref{fig:four_spectrum_dt} further shows changes of the power spectra of $\Bband$ in three-year intervals for GONG and HMI data. It is evident that the signal is not uniformly strong over the entire time period, neither persistently present (i.e., significant); instead, it shows a pattern of rising and waning over cycle 24. The strongest and clearest signal is observed during the solar minimum interval of 2019--2021 with a peak amplitude of $\approx 4\times10^{-4}~$G$^{2}$~nHz$^{-1}$, while no significant signal is observed during 2016--2018. We also find a clear signal during 2010--2015 with a peak amplitude of $\approx 3\times10^{-4}~$G$^{2}$~nHz$^{-1}$. GONG and HMI results show very good agreement.

\subsection{Latitudinal variation of phase}
We calculated the average phase of $\Bavg$ as a function of latitude using a temporal Fourier transform. Phase shifts relative to $67.5^\circ$ latitude were extracted within a narrow band centered at 338~nHz ($\pm$10~nHz) and then averaged. The resulting phase, wrapped to the interval [$-\pi$, $\pi$], varies smoothly between $55^\circ$ and $80^\circ$ latitude (see Fig.~\ref{fig:phase} in the Appendix).
The smooth variation suggests that the excess power at higher latitudes reflects a genuine mode signal rather than stochastic fluctuations.
The phase becomes irregular at lower latitudes, where the mode power is not significant.

\section{Comparison of $\Blos$ and $\Vlos$ } \label{sect:comp}

\begin{figure}
\centering \includegraphics[width=1\linewidth,angle=0]{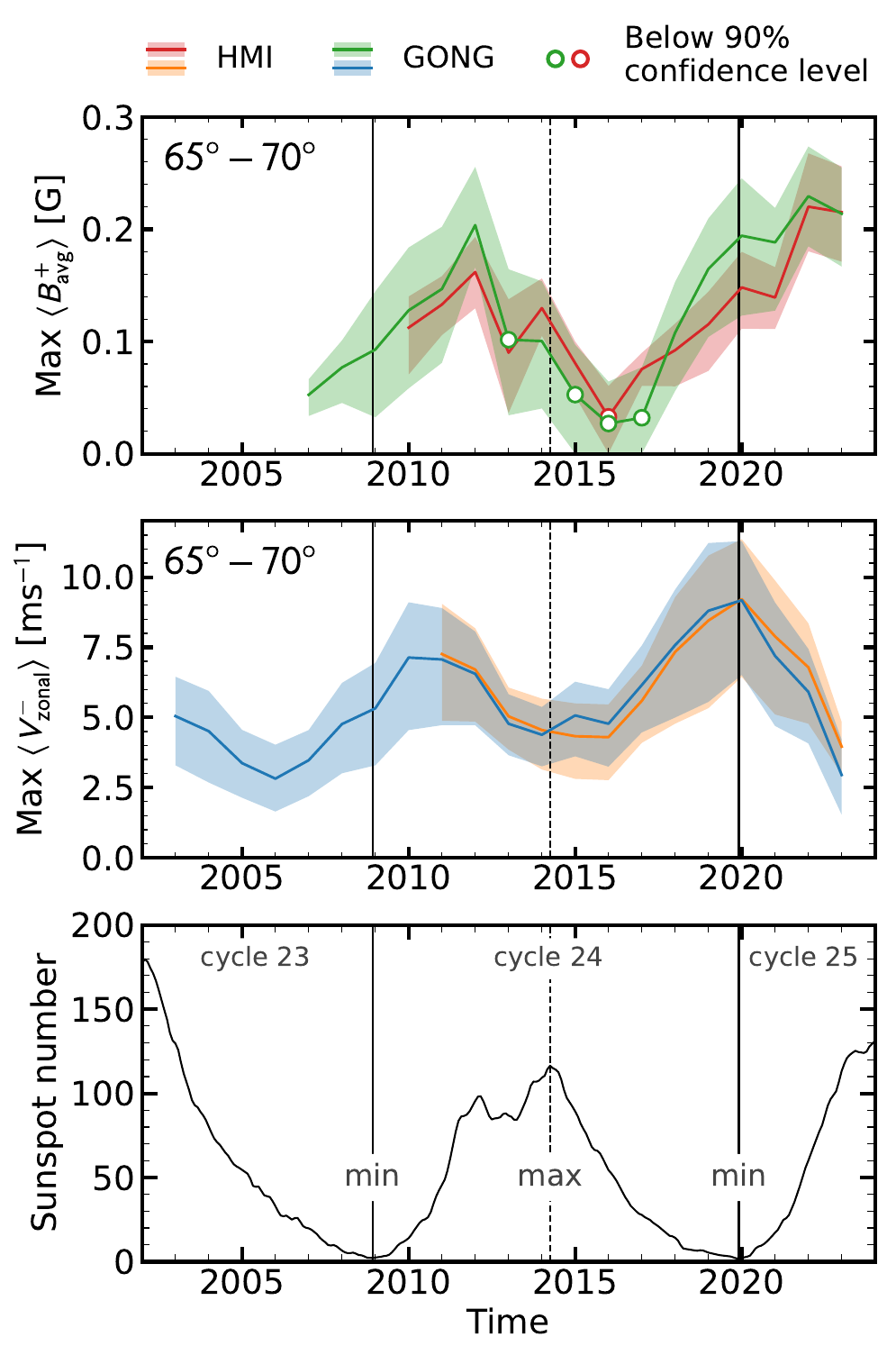}
\caption{Temporal variations of the $m=1$ mode amplitude extracted from  $\Blos$  and $\Vlos$ data. The red, orange, green, and blue curves correspond to results from the HMI $\Bband$, HMI $\Vband$, GONG $\Bband$, and GONG $\Vband$ datasets, respectively. Shaded regions indicate the $68\%$ confidence interval, estimated using 10,000 Monte Carlo simulations. Open circles denote instances where the excess power near the mode frequency falls below the $90\%$ confidence level.
We note that $\Vband$ is derived directly from the LOS Doppler velocity (Eqs.~\eqref{eq:vzonal}--\eqref{eq:vband}); unlike \citetalias{2024Liang_vlosm1}, we did not apply a multiplicative factor in the present work to scale the \(\Vband\) amplitudes to match the horizontal velocities from the ring-diagram maps.
The numerical values used in these plots are available in a supplementary file. Vertical lines denote key phases of the solar cycle, with solid lines representing solar minima and dashed lines indicating solar maxima. The bottom panel shows the average sunspot number (Source: WDC-SILSO, Royal Observatory of Belgium, Brussels, \url{https://doi.org/10.24414/qnza-ac80}).}
\label{fig:cycle}
\end{figure}

We now compare the excess power around 338~nHz in the $\Bband$ and $\Vband$ spectra. Following the procedure described by \citetalias{2024Liang_vlosm1} (their section 5.2), we estimate the total power around the mode frequency by fitting a Lorentzian function. Figure~\ref{fig:cycle} shows the temporal variations of the mode amplitude measured from overlapping 3-yr time series, with central times spaced one year apart, starting in 2007 for $\Bband$ and in 2003 for $\Vband$.

We find very good agreement between the GONG and HMI results for each respective quantity. The mode amplitudes seen in $\Bband$ and $\Vband$ are positively correlated, but there is a time lag between the magnetic field and velocity perturbations. The maximum amplitude in $\Bband$ is about $0.2$~G in 2012 and again around 2020--2023. Both $\Bband$ and $\Vband$ are moderately anti-correlated with the sunspot number, but the mode amplitudes measured in $\Bband$ appears to reach a  maximum during  rising phases of the sunspot cycle.

\begin{figure*}
\sidecaption 
\includegraphics[width=0.7\linewidth]{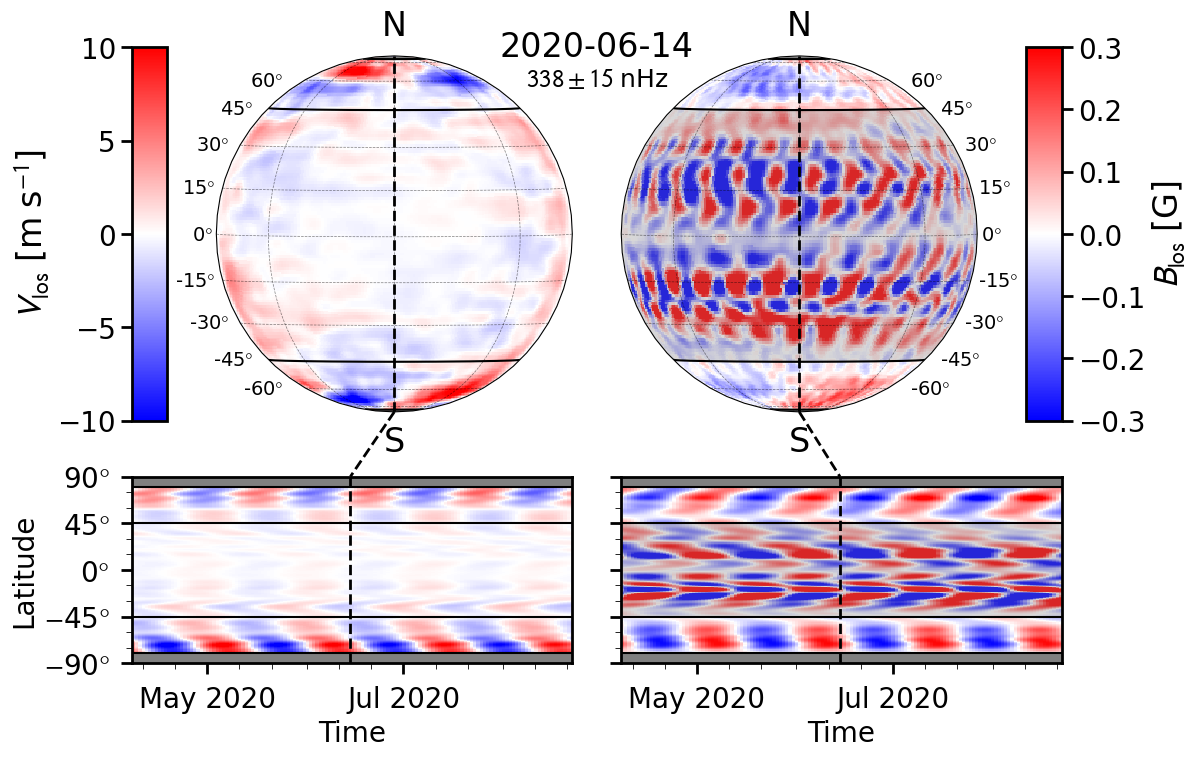}
\caption{Bandpass-filtered images of the LOS Doppler velocity, $\Vlos$, and LOS magnetic field, $\Blos$, derived from HMI data. The top-left panel shows a filtered $\Vlos$ image from 14 June 2020, while the bottom-left panel displays the central-meridian $\Vlos$ values stacked over time and plotted as a function of time and latitude. The right panels display the corresponding $\Blos$ data, with low-latitude regions ($<45^{\circ}$) shaded out. The filtering was performed using a bandpass filter centered at $\nu_{\rm HL1}^{\rm synodic} = 338$~nHz with a full width of 30~nHz, applied over the entire available HMI time period from 2010 to 2024. For clarity, the images were smoothed with a Gaussian kernel with a width of $2$ pixel---in longitude and latitude for the top panels, and in time and latitude for the bottom panels. A movie is available as online ({\color{red} link to A\&A}) supplementary material.}\label{fig:flt_img}
\end{figure*}

Figure~\ref{fig:flt_img} shows frequency-filtered images of the LOS Doppler velocity, $\Vlos$ and the LOS magnetic field $\Blos$ derived from HMI observations. A bandpass filter centered at $\nu_{\rm HL1}^{\rm synodic} = 338$~nHz with a full width of 30~nHz was applied to retain the $m=1$ mode. We emphasize that no spatial Fourier transform was applied. For $\Vlos$, the structure of the mode is clearly visible at high latitudes, while little to no signal is present at lower latitudes. The $\Blos$ data also shows the mode at high latitudes, but the lower latitudes display a plethora of signals associated with residual magnetic activity and not related to the mode. 
The bottom panels of Fig.~\ref{fig:flt_img} show synoptic maps of $\Vlos$ and $\Blos$, constructed by stacking the central meridian from bandpass-filtered images over time. The $m=1$ mode appears as inclined stripes in time-latitude space. A similar pattern was already visible in the synoptic maps of Fig.~\ref{fig:supersynoptic_HL} in the absence of a frequency filter.

\section{Discussion} \label{sect:disc}

We identified a coherent magnetic field oscillation at a frequency of $338$~nHz (with a frequency resolution of $10.5$~nHz) and a peak amplitude of about $0.2$~G in $\Blos$. This oscillation is associated with the high-latitude $m=1$ global inertial mode previously characterized in velocity data. The $\Blos$ perturbations are predominantly symmetric across the equator, in contrast to the $\Vlos$ perturbations, which are primarily north-south antisymmetric. 
The magnetic field oscillations are most clearly observed at latitudes above $60^\circ$ during the recent solar minimum (2018--2022). The mode amplitude in $\Bband$ peaks around 2012, during the rising phase of cycle 24, and again during 2020--2023, the rising phase of cycle 25. A time lag of approximately two years is observed between the trends in magnetic and velocity perturbations.

As outlined in a simple model in Appendix~\ref{app:model}, the amplitude of the magnetic oscillation during solar cycle minimum may be understood via the linearized induction equation, Eq.~\eqref{eq:lin_ind}, whereby the radial magnetic field is advected by the hydrodynamic high-latitude mode. This simple model reproduces approximately the symmetric pattern at high latitudes with a maximum amplitude of $\approx0.2$~G. The calculation assumes a dipolar background field magnetic field during solar minimum (Fig.~\ref{fig:modelB0}). It would be interesting to repeat these calculations over different phases of the solar cycle using, e.g.,  a background magnetic field from a dynamo model.

\begin{acknowledgements}
 The observations were analyzed by SGH and Z-CL. The analytical model was derived by LG and Z-CL.  SGH acknowledges funding from the Austrian Science Fund (FWF) Erwin-Schr\"odinger fellowship J-4560 and funding from the Research Council of Finland (Academy Fellowship): 370747 (RIB-Wind). We thank B.~Proxauf and  S.~Good for valuable discussions. This work utilizes GONG data from NSO, which is operated by AURA under a cooperative agreement with NSF and with additional financial support from NOAA, NASA, and USAF. The HMI data used are courtesy of NASA/SDO and the HMI science team.
\end{acknowledgements}

\bibliographystyle{aa}
\bibliography{biblio.bib}


\begin{appendix}

\onecolumn
\section{Phase of magnetic oscillations as a function of latitude}
\begin{figure*}[h]
\sidecaption
\includegraphics[width=11cm,trim={0 0.32cm 0 0},clip]{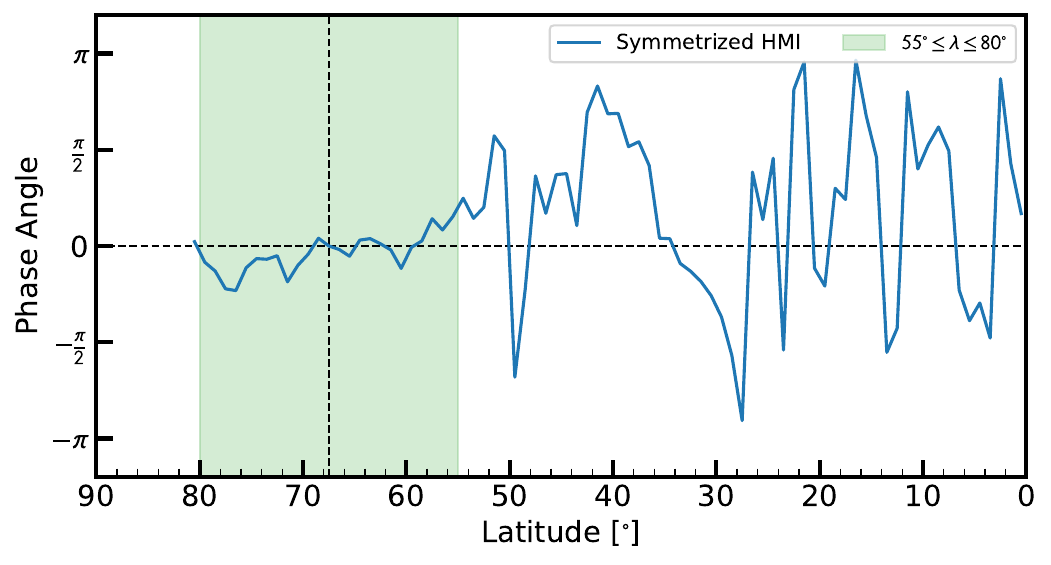}
\caption{
Latitudinal dependence of the phase of $\Bavg$ relative to latitude $67.5^\circ$, measured during solar minimum (2018--2021) and at the frequency of the $m=1$ high-latitude mode.
The green shaded area highlights the latitude range where the phase varies smoothly. This coincides with the latitude range where the mode has the largest amplitudes in the power spectrum.}
\label{fig:phase}
\end{figure*}

\section{Simplified model for magnetic field perturbations during cycle minimum}\label{app:model}

Let us consider the induction equation in an inertial frame,
\begin{equation} \label{eq:induction}
\partial_t \bB = \eta \Delta \bB + \nabla\times(\bv \times \bB) .
\end{equation}
Assuming background values of the magnetic field ($\bB_0$) and of the axisymmetric flow ($\bv_0$) during solar minimum, we wish to compute the perturbations to the magnetic field at the surface ($\bB'$) caused  by the fluctuating flow $\bv'$ associated with a  mode of oscillation.  In the near-surface layers, the velocity field of a quasi-toroidal mode of oscillation is approximately horizontal and divergence free,
\begin{equation}
    \bv' =  \Re \left\{ \bu'(\theta) e^{\ii m \phi  - \ii \omega t} \right\}, \quad \nabla\cdot\bv' = 0.
\end{equation}
The magnetic field is assumed to be purely radial at the surface. We linearize the induction equation to obtain an equation for the magnetic field perturbation  $\bB'$:
\begin{equation} \label{eq:lin_ind}
\partial_t \bB'  - \nabla\times(\bv_0 \times \bB') - \eta \Delta \bB'
= \nabla\times(\bv' \times \bB_0) .
\end{equation}
Taking the radial component of this equation, we have
\begin{align}
\partial_t B_r' + \nabla\cdot (\bv_0 B_r' )  - \eta \Delta  B_r'   =  - ( \bv'\cdot\nabla) B_{0 r} ,
\label{eq.brprime}
\end{align}
where we used $\nabla\cdot \bB_0=0$ and $\nabla\cdot \bv'=0$. 

\begin{figure}[h]
  \resizebox{\hsize}{!}{\includegraphics
  {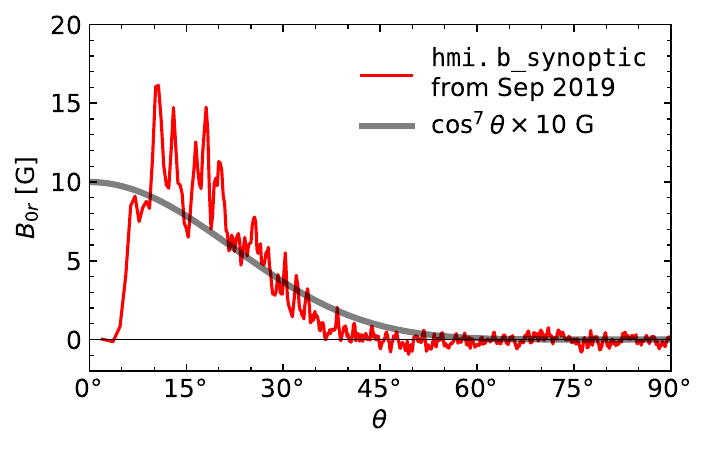}}
  \caption{
  Background radial magnetic field during solar minimum estimated from the data series \texttt{hmi.b\_synoptic} \citep{Liu2017}.
  The synoptic maps for the radial component of the magnetic field were stacked in time and then Gaussian smoothed with a full width  of $27$ days.
  The red curve shows the data in the northern hemisphere on 15 September 2019 when the Sun was quiet and its north pole was tilted toward the Earth. The black curve shows the approximation $B_{0r}(\theta)=10 \times (\cos\theta)^7$~G, which we use in  Eqs.~\eqref{eq:expanded} and~\eqref{eq:simplified}.
  }
  \label{fig:modelB0}
\end{figure}

During solar minimum, the background field  at the surface is approximately radial at the surface and independent of longitude, that is   $\bB_0=B_{0r}(\theta)\ \hat{\br}$. As shown in Fig.~\ref{fig:modelB0}, a reasonable model for the  field is $B_{0r}  \approx b_0  (\cos \theta)^n$ with  $b_0= 10$~G and $n=7$. The background flow consists of  rotation and the meridional flow, that is $\bv_0 = R\sin\theta\ \Omega(\theta)\ \hat{\boldsymbol{\phi}} +  v_{\rm MC}(\theta)\ \hat{\boldsymbol{\theta}}$. We use the surface angular velocity profile  $\Omega(\theta)$ from \cite{Snodgrass1984} and the meridional flow profile $v_{\rm MC}(\theta) = -15\,\sin(2\theta)\,\sin\theta$~m/s.

\begin{figure}[t]
  \resizebox{\hsize}{!}{\includegraphics{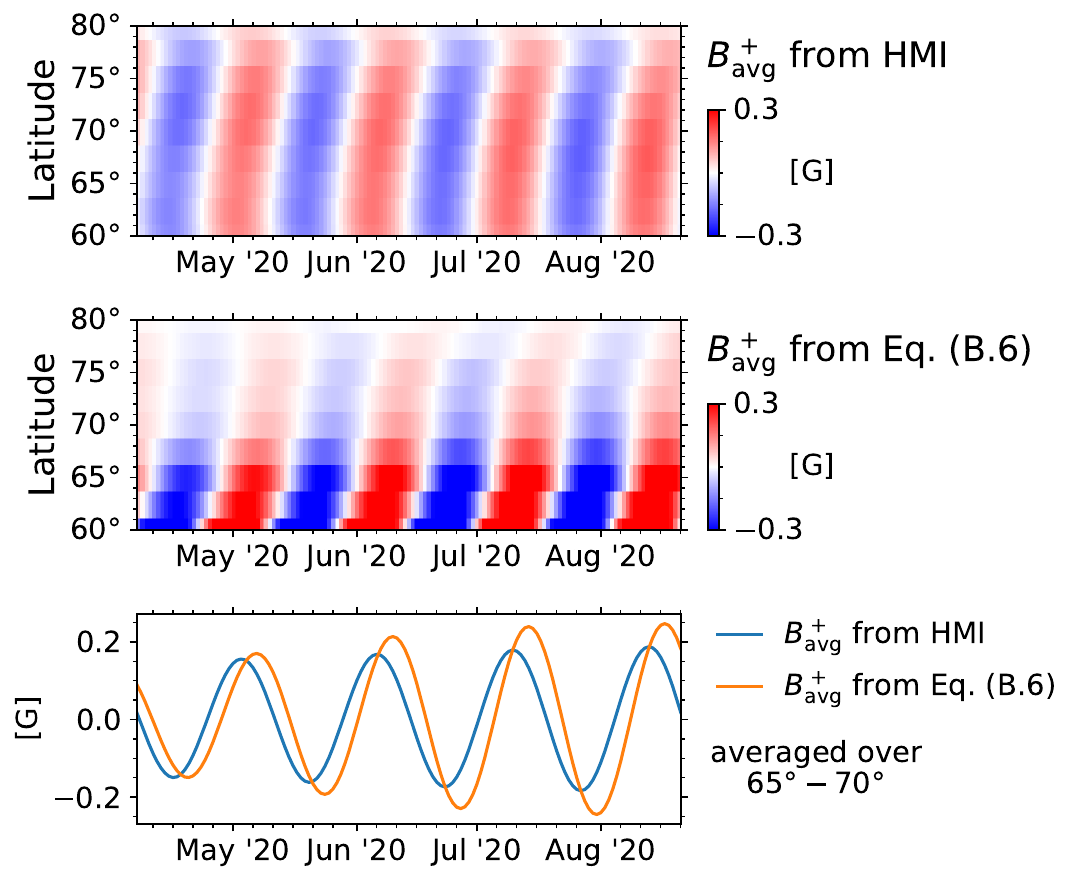}}
  \caption{Top panel: Observed $\Bavg$ using HMI magnetograms at the HL1 mode frequency during solar minimum.
Middle panel: Estimate of $\Bavg$ above the critical latitude using the approximation given by Eq.~\eqref{eq:simplified}. Bottom panel: comparison of observed and modeled values, after averaging over the latitudinal band $65^\circ$--$70^\circ$.} \label{fig:cf_br}
\end{figure}

Next, we plug perturbations to the magnetic field of the form $B_r' = \Re \left\{ b'(\theta) e^{\ii m\phi -\ii \omega t}  \right\}$
into Eq.~(\ref{eq.brprime}) to obtain
\begin{equation}
\label{eq:expanded}
-\ii \omega b' + \ii m \Omega b' + \frac{1}{R\sin\theta}\frac{d}{d\theta}\left( \sin\theta \left( v_{\rm MC}\ b'  - \frac{\eta}{R}\frac{d b'}{d\theta} \right)\right) + \frac{\eta m^2  b'}{ (R\sin\theta)^2 } 
=  - \frac{u'_\theta (\theta)}{R} \frac{d B_{0r}}{ d\theta} .
\end{equation}
We adopt a surface diffusivity of $\eta = 250$~km$^2$/s, consistent with the supergranulation. The RHS of the equation depends on the colatitudinal component of the $m=1$ high-latitude (HL1) mode velocity,
 $u'_\theta(\theta)$, which we extract from HMI ring-diagram flow maps along the central meridian at the HL1 mode frequency.
As a  sanity check, we compute separately the left and side (LHS) of the equation using the observed $b'$ as input, and the right hand side (RHS) of the equation using the observed $u'_\theta$. 
At latitudes above $60^\circ$ (i.e. above the critical latitude of the HL1 mode), we find that the two sides of the equation have comparable amplitudes. We also find that the diffusion and meridional flow terms on the LHS of Eq.~\eqref{eq:expanded} are negligible.
This implies that  a rough estimate for $b'$ is 
\begin{equation} \label{eq:simplified}
   b' (\theta) \approx  \frac{ \ii  u'_\theta  }{R( m \Omega-\omega )}  \frac{d B_{0r}}{ d\theta}    \quad \textrm{for } \theta \ll  \theta_c,
\end{equation}
where   $\theta_c=32^\circ$ is the critical colatitude given by $\Omega(\theta_c)=\omega/m$. Figure~\ref{fig:cf_br} compares the line-of-sight projection of the fluctuating magnetic field derived from Eq.~\eqref{eq:simplified} to the observed  $\Bavg$  during cycle minimum (also see Figure~\ref{fig:supersynoptic_HL}).
Although the approximate model does not perfectly reproduce the observations (esp. the phase), it provides the correct order-of-magnitude estimate for the observed $\Bavg$. We therefore conclude that the observed magnetic field perturbations at the surface are largely  caused by the passive advection of the radial field $B_r'$ by the HL1 mode (velocity $v'_\theta$). 

\twocolumn

\end{appendix}

\end{document}